\title{Deep Historical Borrowing Framework to Prospectively and Simultaneously Synthesize Control Information in Confirmatory Clinical Trials with Multiple Endpoints}
\author{
}

\author{Tianyu Zhan$^{1, \dagger}$, Yiwang Zhou$^2$, Ziqian Geng$^1$, Yihua Gu$^1$, Jian Kang$^3$, \\ Li Wang$^1$, Xiaohong Huang$^1$ and Elizabeth H. Slate$^4$}

\date{ \small
	$^1$ Data and Statistical Sciences, AbbVie Inc., North Chicago, IL, USA \\
	$^2$ Department of Biostatistics, St. Jude Children’s Research Hospital, Memphis, TN, USA \\
	$^3$ Department of Biostatistics, University of Michigan, Ann Arbor, MI, USA \\
	$^4$ Department of Statistics, Florida State University, Tallahassee, FL, USA \\
	$^\dagger$ Corresponding author: Tianyu Zhan, 1 Waukegan Road, North Chicago, IL 60064, USA. {tianyu.zhan.stats@gmail.com} }

\documentclass[12pt]{article}
\usepackage[margin=0.7in]{geometry}

\usepackage{amsmath}
\usepackage{amsfonts}
\usepackage{amssymb}
\usepackage{yhmath}
\usepackage{boldline}
\usepackage{graphicx}
\usepackage{filecontents}
\usepackage{setspace}
\usepackage{amsmath}
\usepackage{algorithm}
\usepackage{algorithmic}
\usepackage{array,booktabs}
\usepackage{natbib}
\usepackage{tikz}
\usepackage{subcaption}
\usepackage{hyperref}
\usetikzlibrary{matrix,chains,positioning,decorations.pathreplacing,arrows}

\def\layersep{2.5cm}

\begin{document}
\maketitle

\section*{\center Abstract}

In current clinical trial development, historical information is receiving more attention as it provides utility beyond sample size calculation. Meta-analytic-predictive (MAP) priors and robust MAP priors have been proposed for prospectively borrowing historical data on a single endpoint. To simultaneously synthesize control information from multiple endpoints in confirmatory clinical trials, we propose to approximate posterior probabilities from a Bayesian hierarchical model and estimate critical values by deep learning to construct pre-specified strategies for hypothesis testing. This feature is important to ensure study integrity by establishing prospective decision functions before the trial conduct. Simulations are performed to show that our method properly controls family-wise error rate (FWER) and preserves power as compared with a typical practice of choosing constant critical values given a subset of null space. Satisfactory performance under prior-data conflict is also demonstrated. We further illustrate our method using a case study in Immunology. 

{\bf Keywords}: Bayesian hierarchical model; Deep learning; Family-wise error rate control; Power preservation; Prospective algorithm 

\section{Introduction}
\label{sec:intro}

Historical control data are usually summarized as estimates of parameters needed to calculate the sample size when designing a traditional Phase III randomized clinical trial \citep{chow2007sample}. This relevant information can be properly borrowed for the current trial to make it more efficient and ethical by allowing fewer patients randomized to the control group or decreasing the total sample size \citep{berry2010bayesian, viele2014use}. Some challenges exist in applying this framework to clinical trials, especially confirmatory studies. While it is always possible to {\it retrospectively} use historical information once the new evidence is available, it is appealing to ensure study integrity by designing a  {\it prospective} algorithm for leveraging historical data \citep{neuenschwander2010summarizing}. Moving beyond the ``sweet spot'' where the borrowed information and the current data are similar, one needs to properly discount historical information to control the bias and the type I error rates \citep{viele2014use}. 

In the context of a single endpoint, \cite{neuenschwander2010summarizing} proposed a novel meta-analytic-predictive (MAP) approach to prospectively borrow historical information for the current trial. \cite{schmidli2014robust} further developed an innovative method to approximate the MAP prior by a mixture of conjugate priors, and therefore the posterior distribution is available in a closed form. A robust MAP prior is then formulated by adding a weakly informative component to discount historical data under prior-data conflict \citep{schmidli2014robust}. Moving further to confirmatory clinical trials, most use multiple endpoints to assess the effects of the study drug \citep{fda2017}. A Bayesian hierarchical model is a natural approach to simultaneously synthesize information from multiple endpoints \citep{berry2010bayesian}. However, taking a trial with binary endpoints as an example, additional non-trivial work is needed to generalize the MAP framework to approximate the joint prior of response rates and to investigate whether the resulting multivariate posterior distribution is available analytically. 

As an alternative, we propose a two-stage Deep Neural Networks (DNN) guided algorithm to build pre-specified decision functions before initiation of the current trial. It takes advantage of the strong functional representation of DNN \citep{goodfellow2016deep, bach2017breaking, yarotsky2017error} to directly approximate posterior probabilities in Section \ref{sec:approx}, and critical values in Section \ref{sec:fwer}. Our proposed method has several appealing features. First, it is a prospective approach in the sense that pre-trained DNN models can be locked in files before initiation of the current trial to ensure study integrity. Moreover, our method provides an accurate type I error rate control by modeling corresponding critical values. As an alternative, simulation-based type I error control by using constant critical values within a subset of the null space can be viewed as a ``worst case scenario'' adjustment, which leads to power loss when the type I error rate is conservative \citep{proschan1995designed, graf2014maximum, zhan2020modified}. More demonstration is provided in Section \ref{sec:sim}. Additionally, simulations and the case study show that our method has relatively small bias and mean squared error (MSE) under prior-data conflict by properly discounting prior information. 

The remainder of this article is organized as follows. In Section \ref{sec:bay}, we introduce a Bayesian hierarchical model on control data from several historical studies. Next we propose DNN-based algorithms to approximate the posterior probabilities and critical values to build pre-specified decision functions for hypothesis testing in Section \ref{sec:DNN_framework}. Simulations in Section \ref{sec:sim} and a case study in Section \ref{sec:case} are conducted to evaluate the performance of our method. Concluding remarks are provided in Section \ref{sec:discussion}.

\section{A Bayesian hierarchical model on historical control data}
\label{sec:bay}

Consider a two-group randomized controlled clinical trial with $I$ ($I \geq 2$) endpoints to study the efficacy of a treatment versus placebo. We consider a setup of $I=2$ binary endpoints for illustration, but our method can be readily generalized to $I>2$ endpoints and other types of endpoints. Denote $R^{(t)}_{i, 0}$ as the number of responders in the current treatment group for endpoint $i$, where $i \in \{1, \cdots, I \}$, and $n^{(t)}_0$ as the total number of subjects from the treatment arm in the current trial. The superscript ``(t)'' denotes the treatment group. For each endpoint $i$, a Beta conjugate prior is assumed on the Binomial sampling distribution with rate $\psi^{(t)}_{i, 0}$,
\begin{equation}
\label{equ:trt_model}
R^{(t)}_{i, 0} \mid \psi^{(t)}_{i, 0} \sim Binomial\left\{ n^{(t)}_{0}, \psi^{(t)}_{i, 0} \right\}, \:\:\: \psi^{(t)}_{i, 0} \sim Beta(a_i, b_i), \:\:\:i = 1, \cdots, I. 
\end{equation}

The control data are available in the current trial and $J$ historical studies. The corresponding notations are denoted as $R^{(c)}_{i, j}$ and $n^{(c)}_j$, where $j=0$ indicates the current trial, $j \in \left\{1, \cdots, J \right\}$ is the index of historical study $j$, and $i \in \left\{1, \cdots, I \right\}$ refers to endpoint $i$. We consider the following Bayesian hierarchical model on the control data \citep{neuenschwander2010summarizing, schmidli2014robust},
\begin{equation}
\label{equ:bay_hie_model}
R^{(c)}_{i,j} \mid \psi^{(c)}_{i,j} \sim Binomial\left\{ n^{(c)}_{j}, \psi^{(c)}_{i,j}\right\}, \:\:\: \mu_{i,j} = logit\left\{ \psi^{(c)}_{i,j} \right\}, \:\:\boldsymbol{\mu}_j \sim MVN (\boldsymbol{\theta}, \Sigma),
\end{equation}
where $\boldsymbol{\mu}_j = (\mu_{1,j}, \cdots, \mu_{I,j})$, for $i \in \left\{1, \cdots, I \right\}$, $j \in \left\{0, 1, \cdots, J \right\}$, and $ MVN (\boldsymbol{\theta}, \Sigma)$ denotes a multivariate Normal distribution with mean vector $\boldsymbol{\theta}$ and variance-covariance matrix $\Sigma$. A vague prior is assumed on $\boldsymbol{\theta}$, and an $InverseWishart(\Sigma_0, k)$ prior is assigned to the variance-covariance matrix $\Sigma$ with positive definite $I \times I$ matrix $\Sigma_0$ and degrees of freedom $k \geq I$. The expectation of a $Wishart(\Sigma_0, k)$ is $k \left(\Sigma_0\right)^{-1}$, and therefore $ \Sigma_0/k$ is a prior guess for $\Sigma$. We use $\boldsymbol{D}_H = \left[R^{(c)}_{i,j}, n^{(c)}_{j}, i \in \left\{1, \cdots, I \right\}, j \in \left\{1, \cdots, J \right\} \right]$ to denote control information in $J$ historical studies, and $\boldsymbol{D}_N = \left[R^{(c)}_{i,0}, n^{(c)}_{0}, R^{(t)}_{i,0}, n^{(t)}_{0}, i \in \left\{1, \cdots, I \right\} \right]$ as data in the current new trial. 

Our quantity of interest is the posterior probability of observing a promising treatment effect in the current trial,  
\begin{equation}
\label{equ:bay_interest}
S_i = Pr \left\{ \psi_{i, 0}^{(t)} - \psi^{(c)}_{i, 0}> \theta_i \middle| \boldsymbol{D}_H, \boldsymbol{D}_N \right\}, \:\:\: i = 1, \cdots, I,
\end{equation}
where $\theta_i$ is a pre-specified constant for endpoint $i$. The decision function of rejecting the null hypothesis pertaining to endpoint $i$ is if $S_i > \widetilde{c}_i$. The critical value $\widetilde{c}_i$ can interpreted as a threshold to claim a significant treatment effect in the $i$th endpoint with controlled type I error rates. The computation of $\widetilde{c}_i$ is studied in the next section to control the family-wise error rate (FWER) at a nominal level $\alpha$. We denote $\boldsymbol{S} = (S_1, \cdots, S_I)$ as a vector of those posterior probabilities. 

Since the posterior probabilities $\boldsymbol{S}$ in (\ref{equ:bay_interest}) usually do not have closed forms, we estimate them empirically by a large number of Monte Carlo samples. Specifically, posterior samples $\psi_{i, 0}^{(t)}$ are simulated from the conjugate prior model in (\ref{equ:trt_model}), and posterior samples $\psi_{i, 0}^{(c)}$ in the Bayesian hierarchical model (\ref{equ:bay_hie_model}) are obtained by the Markov chain Monte Carlo (MCMC) method based on current trial data \citep{berry2010bayesian}. However, it is appealing to build a prospective algorithm before conducting the current new trial to ensure the study integrity. In studies with a single binary endpoint ($I=1$), \citet{schmidli2014robust} proposed a novel approach by approximating the Meta Analytic Predictive (MAP) prior $p\left\{\psi^{(c)}_{i, 0} | \boldsymbol{D}_H\right\}$ with a mixture of Beta distributions, and hence the posterior distribution $p\left\{ \psi^{(c)}_{i, 0} | \boldsymbol{D}_H, \boldsymbol{D}_N\right\}$ becomes a weighted average of Beta distributions. In the context of multiple endpoints ($I \geq 2$), a Bayesian hierarchical model in (\ref{equ:bay_hie_model}) is a natural approach to simultaneously synthesize control information \citep{berry2010bayesian}. Even if one can utilize several base distributions to approximate the multivariate prior $p\left\{ \boldsymbol{\psi}_0^{(c)} | \boldsymbol{D}_H \right\}$, where $\boldsymbol{\psi}_0^{(c)} = \left\{ \psi^{(c)}_{1, 0}, \cdots, \psi^{(c)}_{I, 0} \right\}$, the joint posterior probability $\boldsymbol{S}$ does not necessarily have an analytic closed form. 

As an alternative, we propose to directly approximate $\boldsymbol{S}$ based on observed historical data $\boldsymbol{D}_H$ and varying simulated new trial data $\boldsymbol{D}_N$ by deep neural networks (DNN) in the study design stage. After collecting results from the current trial, one can instantly compute $\boldsymbol{S}$ and conduct downstream hypothesis testing based on pre-specified approximation functions.

\section{A DNN-based historical borrowing framework}
\label{sec:DNN_framework}

In this section, we first provide a short review on DNN in Section \ref{sec:review}, and then introduce our DNN guided historical borrowing framework by directly approximating the posterior probabilities and posterior means in Section \ref{sec:approx}. In Section \ref{sec:fwer}, we further estimate corresponding critical values by DNNs to control FWER in the strong sense. The hypothesis testing based on observed current trial data is discussed in Section \ref{sec:hypo}. 

\subsection{Review of DNN}
\label{sec:review}

\begin{figure}
	\centering
	\begin{tikzpicture}[shorten >=1pt,->,draw=black!50, node distance=\layersep]
	\tikzstyle{every pin edge}=[<-,shorten <=1pt]
	\tikzstyle{neuron}=[circle,fill=black!25,minimum size=17pt,inner sep=0pt]
	\tikzstyle{input neuron}=[neuron, fill=green!50];
	\tikzstyle{output neuron}=[neuron, fill=red!50];
	\tikzstyle{hidden neuron}=[neuron, fill=blue!50];
	\tikzstyle{annot} = [text width=4em, text centered]
	
	\foreach \name / \y in {1,...,4}
	\node[input neuron, pin=left:Input \#\y] (I-\name) at (0,-\y-0.5) {};
	
	\foreach \name / \y in {1,...,6}
	\path[yshift=0.5cm]
	node[hidden neuron] (H-\name) at (\layersep,-\y cm) {};
	
	\foreach \name / \y in {1,...,6}
	\path[yshift=0.5cm]
	node[hidden neuron] (L-\name) at (\layersep+\layersep,-\y cm) {};
	
	\node[output neuron,pin={[pin edge={->}]right:Output \#1}, right of=L-3] (O1) {};
	\node[output neuron,pin={[pin edge={->}]right:Output \#2}, right of=L-4] (O2) {};
	
	\foreach \source in {1,...,4}
	\foreach \dest in {1,...,6}
	\path (I-\source) edge (H-\dest);
	
	\foreach \source in {1,...,6}
	\foreach \dest in {1,...,6}
	\path (H-\source) edge (L-\dest);
	
	\foreach \source in {1,...,6}
	\path (L-\source) edge (O1);
	\foreach \source in {1,...,6}
	\path (L-\source) edge (O2);
	
	\node[annot,above of=H-1, node distance=1cm] (ll) {Hidden layer 1};
	\node[annot,above of=L-1, node distance=1cm] (hl) {Hidden layer 2};
	\node[annot,left of=ll] {Input layer};
	\node[annot,right of=hl] {Output layer};
	\end{tikzpicture}
	\caption{A Deep Neural Network with two hidden layers.}
	\label{F:dnn_example}
\end{figure}
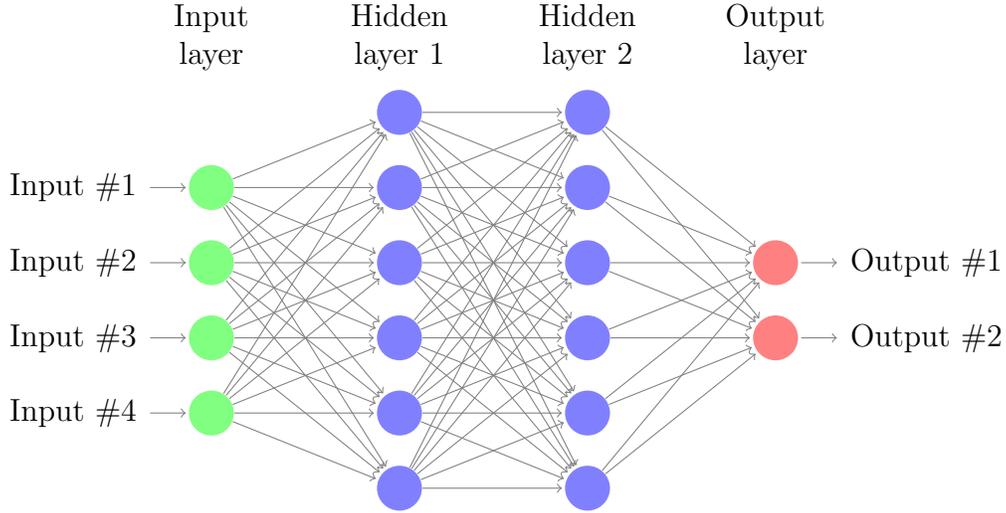

Deep learning is a specific subfield of machine learning as a new take on learning representations from data with successive layers \citep{Chollet}. A major application of Deep Neural Networks (DNN) is to approximate some functions with input data \citep{goodfellow2016deep}. DNN defines a mapping function $F(\boldsymbol{M}; {\boldsymbol{\phi}})$ that learns the value of parameters $\boldsymbol{\phi}$ that result in the best function approximation of output ${\boldsymbol{S}}$ based on input data $\boldsymbol{M}$, where $\boldsymbol{\phi}$ denotes a stack of all weights and bias parameters in the DNN. To simply notations, we use ${\boldsymbol{S}}$ in (\ref{equ:bay_interest}) to denote the output, because later on DNN is utilized to approximate this posterior probability. For example in Figure \ref{F:dnn_example}, the left input $\boldsymbol{M}$ of dimension $4$ is transfered by $2$ hidden layers to approximate a $2$-dimensional output $\boldsymbol{S}$ on the right. 

Typically, training a DNN involves the following four components: layers, input data and corresponding output, loss function, and optimizer \citep{Chollet}. To avoid the potential over-fitting, cross-validation is commonly used to select the architecture from a pool of candidates \citep{goodfellow2016deep}. The loss function measures how well the fitted DNN $F(\boldsymbol{M}; \widehat{\boldsymbol{\phi}})$ approximates the output $\boldsymbol{S}$. The mean squared error (MSE) loss can be utilized if $\boldsymbol{S}$ is continuous. The optimizer determines how the network will be updated based on the loss function. It usually implements a specific variant of the stochastic gradient descent (SGD) algorithm; for example, RMSProp \citep{hinton2012neural} has been shown to be an effective and practical optimization algorithm for DNN \citep{goodfellow2016deep}, and is used in this article. 

\subsection{Posterior probabilities approximation}
\label{sec:approx}

We denote $\boldsymbol{R}^{(c)}_i = \left\{ R^{(c)}_{i,1}, \cdots, R^{(c)}_{i,J} \right\}$ as a stack of numbers of responders in all historical studies $J$ for endpoint $i$, $i = 1, \cdots, I$, and further denote $\boldsymbol{R}_H^{(c)} = \left\{ \boldsymbol{R}^{(c)}_1, \cdots, \boldsymbol{R}^{(c)}_I \right\}$. Corresponding notations are $\boldsymbol{R}^{(c)}_N = \left\{ R^{(c)}_{1,0}, \cdots, R^{(c)}_{I,0} \right\}$ for the current control group, and $\boldsymbol{R}^{(t)}_N = \left\{ R^{(t)}_{1,0}, \cdots, R^{(t)}_{I,0} \right\}$ for the current treatment arm. The subscript ``H'' refers to historical data, while ``N'' corresponds to new trial data. We consider $\mathcal{P}_i^{(c)}$ as a parameter space covering $\psi_{i, j}^{(c)}$ for endpoint $i$ in all $J$ historical studies and $\psi_{i, 0}^{(c)}$ in the current study. For example, $\mathcal{P}_i^{(c)} = \left\{\psi_i^{(c)}: \psi_i^{(c)} \in (0.1, 0.5) \right\}$ indicates that the control response rate for endpoint $i$ in all historical studies and the current study ranges from $0.1$ to $0.5$. It can be set wider as needed. Similarly, $\mathcal{T}_i$ is the parameter space of the treatment effect $\Delta_i$ for endpoint $i$. 

In Algorithm \ref{alg_approx}, we utilize DNN to construct a mapping function $F_S(\boldsymbol{M}; \widehat{\boldsymbol{\phi}})$ to approximate $\boldsymbol{S}$ in (\ref{equ:bay_interest}) based on input data $\boldsymbol{M}$, where $\boldsymbol{M} = \left\{\boldsymbol{R}^{(c)}_N, \boldsymbol{R}^{(t)}_N\right\}$ and $\widehat{\boldsymbol{\phi}}$ are the estimated parameters in DNN. Note that DNN is utilized to estimate the quantity $\boldsymbol{S}$ instead of the posterior distribution. We consider a setup where sample size $\left\{n_0^{(c)}, n_1^{(c)}, \cdots, n_J^{(c)}, n_0^{(t)}  \right\}$ and historical data $\boldsymbol{R}_H^{(c)}$ are constants. Therefore, the simulated input data $\boldsymbol{M} = \left\{\boldsymbol{R}^{(c)}_N, \boldsymbol{R}^{(t)}_N\right\}$ for DNN only contains the number of responders in the current trial. When there are $I=2$ endpoints, $\boldsymbol{M}$ has $4$ elements and $\boldsymbol{S}$ has $2$ elements as shown in Figure \ref{F:dnn_example}. One can build a more general DNN function to accommodate varying sample size and varying historical data. In Step 1, $\boldsymbol{M}$ and $\boldsymbol{R}$ are based on simulated current trial data before the current trial conduct. Their counterparts $\widetilde{\boldsymbol{M}}$ and $\widetilde{\boldsymbol{R}}$ from observed current trial data are plugged in the trained $F_S(\boldsymbol{M}; \widehat{\boldsymbol{\phi}})$ to estimate $\boldsymbol{S}$ at Section \ref{sec:hypo}. 

In Step 2, we perform cross-validation with $80\%$ as the training data and the remaining $20\%$ as the validation data to select a proper DNN structure and other hyperparameters \citep{goodfellow2016deep}. By increasing the number of hidden layers $H_l$, the number of nodes $H_n$ in DNN and the number of training epochs $H_e$, the empirical MSE from the training dataset usually decreases, but the validation MSE may increase. We then apply regulation approaches to increase the generalizability of the model while keeping the training MSE below a certain tolerance, say $10^{-3}$. The regulation approaches include the dropout technique which randomly sets a number of nodes as zeros during training with a dropout rate $H_d$, and the mini-batch approach which stochastically selects a small batch of data with batch size $H_b$ in computing the gradient in the optimization algorithm. Several structures around this sub-optimal structure are added to the candidate pool for cross-validation \citep{goodfellow2016deep, zhan2020optimizing}. The final DNN structure is selected as the one with the smallest validation error and is utilized in Step 3 to obtain the estimated posterior probability $\widehat{\boldsymbol{S}}$ by $\widehat{\boldsymbol{S}} = F_S(\boldsymbol{M}; \widehat{\boldsymbol{\phi}})$. This selected structure is also used for DNN $F_P$ introduced next and for DNNs to estimate critical values in the next section. One can implement this structure selection process when training other DNNs. Since output $\boldsymbol{S}$ is continuous, we use the MSE loss in training the DNN. The estimated posterior probability $\widehat{\boldsymbol{S}}$ is further truncated within $0$ and $1$. The same loss function is used for approximating posterior means of control response rates and for approximating critical values discussed later. On the implementation of DNN, interested readers can refer to \cite{Chollet} and our shared R code with link provided in the Supplementary Materials for more details. 

Similarly, we train another DNN $F_P\left[\boldsymbol{R}^{(c)}_{N}; \widehat{\boldsymbol{\phi}}_P\right]$ to approximate the posterior means of control response rates $\left[\psi_{1, 0}^{(c)}, \cdots, \psi_{I, 0}^{(c)} \right]$. Only the simulated numbers of control responders $\boldsymbol{R}^{(c)}_{N}$ are included as input data for DNN $F_P$, because the treatment and control group are assumed to be independent by models (\ref{equ:trt_model}) and (\ref{equ:bay_hie_model}). Other training details are the same with the DNN $F_S$ introduced in the previous paragraph. 

Here we provide some remarks on accommodating correlations between endpoints in the Step 1 of simulating training data. Per model (\ref{equ:bay_hie_model}), given response rate $\psi_{i, j}^{(c)}$, the number of responders $R_{i, j}^{(c)}$ are independent between endpoints. The correlation is captured by the variance-covariance matrix $\Sigma$ in the multivariate Normal distribution of $\boldsymbol{\mu}_j$. In Step 1, even though we uniformly draw $\psi_{i, 0}^{(c)}$ from its support $\mathcal{P}_i^{(c)}$, we can capture a broad range of correlations on response rates by simulating a sufficiently large number $B$ of training data. In Step 3, the DNN $F_S$ learns the functional form of mapping $\boldsymbol{M}$ to $\boldsymbol{S}$ with correlations accounted. Therefore, we do not directly simulate correlated endpoints in Step 1, but cover varying magnitudes of correlations on response rates in DNN training. In practice, one should check validation error and operating characteristics in testing to make sure that the size $B$ is large enough. An alternative but more complicated approach is to simulate correlated rates $\psi_{i, j}^{(c)}$ with several varying correlation coefficients to reduce training data size $B$. 

\begin{algorithm}
	\caption{Train a DNN $F_S(\boldsymbol{M}; \widehat{\boldsymbol{\phi}})$ to approximate the posterior distributions $\boldsymbol{S}$ based on simulated $\boldsymbol{M}$}
	\label{alg_approx}
	\begin{algorithmic}
\STATE 1. Construct a training dataset for the DNN of size $B$. In each training data $b$, uniformly draw $\psi_{i, 0}^{(c)}$ from $\mathcal{P}_i^{(c)}$, $\Delta_{i, 0}$ from $\mathcal{T}_i$ and set $\psi_{i, 0}^{(t)} = \psi_{i, 0}^{(c)} + \Delta_{i, 0}$, for $i = 1, \cdots, I$. The training input data $\boldsymbol{M} = \left[ \boldsymbol{R}^{(c)}_{N}, \boldsymbol{R}^{(t)}_{N} \right]$ is further simulated from Binomial distributions with their corresponding response rates. The output $\boldsymbol{S}$ in (\ref{equ:bay_interest}) is computed based on simulated $\boldsymbol{M}$, fixed sample sizes $\left\{n_0^{(c)}, n_1^{(c)}, \cdots, n_J^{(c)}, n_0^{(t)}  \right\}$ and observed historical data $\boldsymbol{R}_H^{(c)}$ as demonstrated in Section \ref{sec:bay}. 
\bigskip
\STATE 2. Perform cross-validation on several candidate DNN structures to select one with the smallest validation error for final training.
\bigskip
\STATE 3. Train a DNN to build an approximating function $\widehat{\boldsymbol{S}} = F_S(\boldsymbol{M}; \widehat{\boldsymbol{\phi}})$ to estimate $\boldsymbol{S}$ based on simulated $\boldsymbol{M}$, where $\widehat{\boldsymbol{\phi}}$ are the estimated parameters in DNN. 
	\end{algorithmic}
\end{algorithm}

\subsection{FWER control in the strong sense}
\label{sec:fwer}

In this section, we discuss how to compute the critical value $\widetilde{c}_i$ in the decision rule $S_i > \widetilde{c}_i$ of rejecting the null hypothesis pertaining to endpoint $i$ to strongly control FWER at a nominal level $\alpha$. 

Family-wise error rate (FWER) is the probability of rejecting at least one true null hypothesis. FWER is said to be controlled at level $\alpha$ in the strong sense if it does not exceed $\alpha$ under any configuration of true and false hypotheses \citep{bretz2016multiple}. Define $H_1$ and $H_2$ as the single null hypotheses where only endpoint $1$ or $2$ has no treatment effect, and $H_{12}$ as the global null hypothesis where neither endpoint has treatment effects. In the context of $I=2$ endpoints, we need to control the following three erroneous probabilities,
\begin{align}
& Pr\Big\{ S_1 > c_1 \Big| H_1 \Big\} \leq \alpha \label{def:error_H1}, && H_1: \psi_{1, 0}^{(c)} = \psi_{1, 0}^{(t)}, \psi_{2, 0}^{(c)} < \psi_{2, 0}^{(t)} \\
& Pr\Big\{ S_2 > {c}_2 \Big| H_2 \Big\} \leq \alpha \label{def:error_H2}, && H_2: \psi_{2, 0}^{(c)} = \psi_{2, 0}^{(t)}, \psi_{1, 0}^{(c)} < \psi_{1, 0}^{(t)}\\
& Pr\Big\{ \left(S_1 > {c}_{12}\right) \cup \left(S_2 > {c}_{12} \right) \Big| H_{12} \Big\} \leq \alpha, \label{def:error_H12} && H_{12}: \psi_{1, 0}^{(c)} = \psi_{1, 0}^{(t)}, \psi_{2, 0}^{(c)} = \psi_{2, 0}^{(t)}
\end{align}
where ${c}_1$ is the critical value to control error rate under $H_1$, ${c}_2$ for $H_2$, and ${c}_{12}$ for $H_{12}$. 

In Algorithm \ref{alg_critical}, we train three DNNs to estimate these three critical values: $c_1$, $c_2$ and $c_{12}$. Taking $H_1$ in (\ref{def:error_H1}) as an example, we define $\psi_{1, 0}^{(c, t)}$ as the common value of $\psi_{1, 0}^{(c)}$ and $\psi_{1, 0}^{(t)}$ under $H_1$. The training input data is denoted as $\boldsymbol{M}_1 = \left\{\psi_{1, 0}^{(c, t)}, \psi_{2, 0}^{(c)}, \Delta_{2, 0}\right\}$, which is simulated before the current trial conduct. In Step 1, we simulate $B_1$ varying $\boldsymbol{M}_1$'s from parameter spaces to get the training input data. Given each training feature $\boldsymbol{M}_1$, we then simulate $B_1^\prime$ samples under $H_1$ and compute their estimated posterior probabilities at $\widehat{\boldsymbol{S}} = F_S(\boldsymbol{M}; \widehat{\boldsymbol{\phi}})$ based on the DNN $F_S$ obtained from Algorithm \ref{alg_approx}. The critical value $c_1$ is empirically calculated as the upper $\alpha$ quantile of $\widehat{S}_1$ in $\widehat{\boldsymbol{S}}$ to satisfy (\ref{def:error_H1}). We further train a DNN to obtain a mapping function $\widehat{c}_1 = F_1(\boldsymbol{M}_1; \widehat{\boldsymbol{\phi}}_1)$ to approximate $c_1$ based on simulated $\boldsymbol{M}_1$. Step 2 constructs $\widehat{c}_2 = F_2(\boldsymbol{M}_2; \widehat{\boldsymbol{\phi}}_2)$ under $H_2$ in (\ref{def:error_H2}), and Step 3 computes $\widehat{c}_{12} = F_{12}(\boldsymbol{M}_{12}; \widehat{\boldsymbol{\phi}}_{12})$ under $H_{12}$ in (\ref{def:error_H12}). Similar to Algorithm 1 at Section \ref{sec:approx}, Algorithm 2 is also pre-specified before the current trial conduct in the sense that $\boldsymbol{M}_1$, $\boldsymbol{M}_2$, $\boldsymbol{M}_{12}$ are simulated response rates. Other training details are the same with Section \ref{sec:approx}. 

\subsection{Hypothesis testing based on observed current trial data}
\label{sec:hypo}

After conducting Algorithm 1 in Section \ref{sec:approx} and Algorithm 2 in Section \ref{sec:fwer} based on simulated current trial data, one can save well-trained DNNs in files to ensure the integrity of the current trial conduct. In this section, we illustrate how to perform hypothesis testing based on $\widetilde{\boldsymbol{R}}^{(c)}_N = \left\{\widetilde{R}^{(c)}_{1,0}, \widetilde{R}^{(c)}_{2,0} \right\}$ as the observed number of responder in the current control group, and $\widetilde{\boldsymbol{R}}^{(t)}_N = \left\{\widetilde{R}^{(t)}_{1,0}, \widetilde{R}^{(t)}_{2,0} \right\}$ in the current treatment group. 

The first step is estimate posterior probabilities $\widehat{\boldsymbol{S}} = F_S(\widetilde{\boldsymbol{M}}; \widehat{\boldsymbol{\phi}})$ with the DNN $F_S$ obtained in Algorithm 1, where $\widetilde{\boldsymbol{M}} = \left\{ \widetilde{\boldsymbol{R}}^{(c)}_N, \widetilde{\boldsymbol{R}}^{(t)}_N \right\}$. The next step is to calculate critical values based on three trained DNNs obtained in Algorithm 2. Taking $H_1$ as an example, we denote $\widetilde{\boldsymbol{M}}_1 = \Big[\left\{\widetilde{R}^{(c)}_{1,0}+\widetilde{R}^{(t)}_{1,0}\right\}/\left\{n_0^{(c)}+n_0^{(t)}\right\}, \widetilde{R}^{(c)}_{2,0}/n_0^{(c)},$ $\widetilde{R}^{(t)}_{2,0}/n_0^{(t)} - \widetilde{R}^{(c)}_{2,0}/n_0^{(c)} \Big]$ as the empirical estimator of $\boldsymbol{M}_1 = \left\{\psi_{1, 0}^{(c, t)}, \psi_{2, 0}^{(c)}, \Delta_{2, 0}  \right\}$. We then estimate its critical value at $\widehat{c}_1 = F_1(\widetilde{\boldsymbol{M}}_1; \widehat{\boldsymbol{\phi}}_1)$ based on the trained DNN $F_1$. Correspondingly, we estimate $c_2$ under $H_2$ at $\widehat{c}_2 = F_2(\widetilde{\boldsymbol{M}}_2; \widehat{\boldsymbol{\phi}}_2)$, where
\newline 
$\widetilde{\boldsymbol{M}}_2 = \left[\widetilde{R}^{(c)}_{1,0}/n_0^{(c)}, \widetilde{R}^{(t)}_{1,0}/n_0^{(t)} - \widetilde{R}^{(c)}_{1,0}/n_0^{(c)}, \left\{\widetilde{R}^{(c)}_{2,0}+\widetilde{R}^{(t)}_{2,0}\right\}/(n_0^{(c)}+n_0^{(t)}) \right]$, and estimate $c_{12}$ under $H_{12}$ at $\widehat{c}_{12} = F_{12}(\widetilde{\boldsymbol{M}}_{12}; \widehat{\boldsymbol{\phi}}_{12})$, where $\widetilde{\boldsymbol{M}}_{12} = \left[\left\{\widetilde{R}^{(c)}_{1,0}+\widetilde{R}^{(t)}_{1,0}\right\}/(n_0^{(c)}+n_0^{(t)}), \left\{\widetilde{R}^{(c)}_{2,0}+\widetilde{R}^{(t)}_{2,0}\right\}/(n_0^{(c)}+n_0^{(t)}) \right]$. 

Now we are ready to perform hypothesis testing on treatment effects based on estimated posterior probabilities $\widehat{\boldsymbol{S}} = \left\{ \widehat{S}_1, \widehat{S}_2 \right\}$, and estimated critical values $\widehat{c}_{1}$, $\widehat{c}_{2}$ and $\widehat{c}_{12}$. Since the strong control of FWER is under all configurations of true and false null hypotheses specified in (\ref{def:error_H1}), (\ref{def:error_H2}) and (\ref{def:error_H12}), we set $\widetilde{c}_1 = \max(\widehat{c}_{1}, \widehat{c}_{12})$ for rejecting null hypothesis $H_1$ with the decision function $\widehat{S}_1 > \widetilde{c}_1$, and correspondingly $\widetilde{c}_2 = \max(\widehat{c}_{2}, \widehat{c}_{12})$ for $H_2$. This is analogous to the closure principle of handling multiplicity issues, where the rejection of a particular elementary hypothesis requires the rejection of all intersection hypotheses containing it \citep{tamhane2018advances}. 

\begin{algorithm}
	\caption{Train three DNNs to approximate critical values based on simulated $\boldsymbol{M}_1$, $\boldsymbol{M}_2$ and $\boldsymbol{M}_{12}$}
	\label{alg_critical}
	\begin{algorithmic}
\bigskip
\STATE 1. Under $H_1$ in (\ref{def:error_H1}), simulate training input data $\boldsymbol{M}_1 = \left\{\psi_{1, 0}^{(c, t)}, \psi_{2, 0}^{(c)}, \Delta_{2, 0}  \right\}$ under $H_1$ of size $B_1$. Given each training response rate $\boldsymbol{M}_1$, simulate $B_1^\prime$ sets of responders $\boldsymbol{M}$ and compute their estimated posterior probabilities at $\widehat{\boldsymbol{S}} = F_S(\boldsymbol{M}; \widehat{\boldsymbol{\phi}})$ based on the DNN $F_S$ from Algorithm \ref{alg_approx}. The critical value $c_1$ is computed as the upper $\alpha$ quantile of $\widehat{S}_1$ in $\widehat{\boldsymbol{S}}$. Train a DNN to obtain the mapping function $\widehat{c}_1 = F_1(\boldsymbol{M}_1; \widehat{\boldsymbol{\phi}}_1)$.
\bigskip
\STATE 2. Under $H_2$ in (\ref{def:error_H2}), train another DNN $\widehat{c}_2 = F_2(\boldsymbol{M}_2; \widehat{\boldsymbol{\phi}}_2)$ to estimate $c_2$ based on $\boldsymbol{M}_2 = \left\{ \psi_{1, 0}^{(c)}, \Delta_{1, 0}, \psi_{2, 0}^{(c, t)}\right\}$ of training data size $B_2$ and null data size $B_2^\prime$. 
\bigskip
\STATE 3. Under $H_{12}$ in (\ref{def:error_H12}), the training input data is $\boldsymbol{M}_{12} = \left\{ \psi_{1, 0}^{(c, t)}, \psi_{2, 0}^{(c, t)}\right\}$. The critical value $c_{12}$ is computed by solving a non-linear equation in (\ref{def:error_H12}) based on $\widehat{{S}}_1$ and $\widehat{{S}}_2$ of size $B_{12}^\prime$. The fitted DNN is denoted as $\widehat{c}_{12} = F_{12}(\boldsymbol{M}_{12}; \widehat{\boldsymbol{\phi}}_{12})$.
	\end{algorithmic}
\end{algorithm}

\begin{figure}
	\centering
	\caption{Flowchart of Algorithm 1 and 2 and hypothesis testing. Algorithm 1 utilizes DNN to approximate posterior probability $S$ in (\ref{equ:bay_interest}) based on simulated current trial data. Algorithm 2 trains DNN to estimate critical values $c$ to control type I error rates based on simulated data from null hypothesis. The hypothesis testing is based on observed current trial data. }
	\includegraphics[scale=0.7]{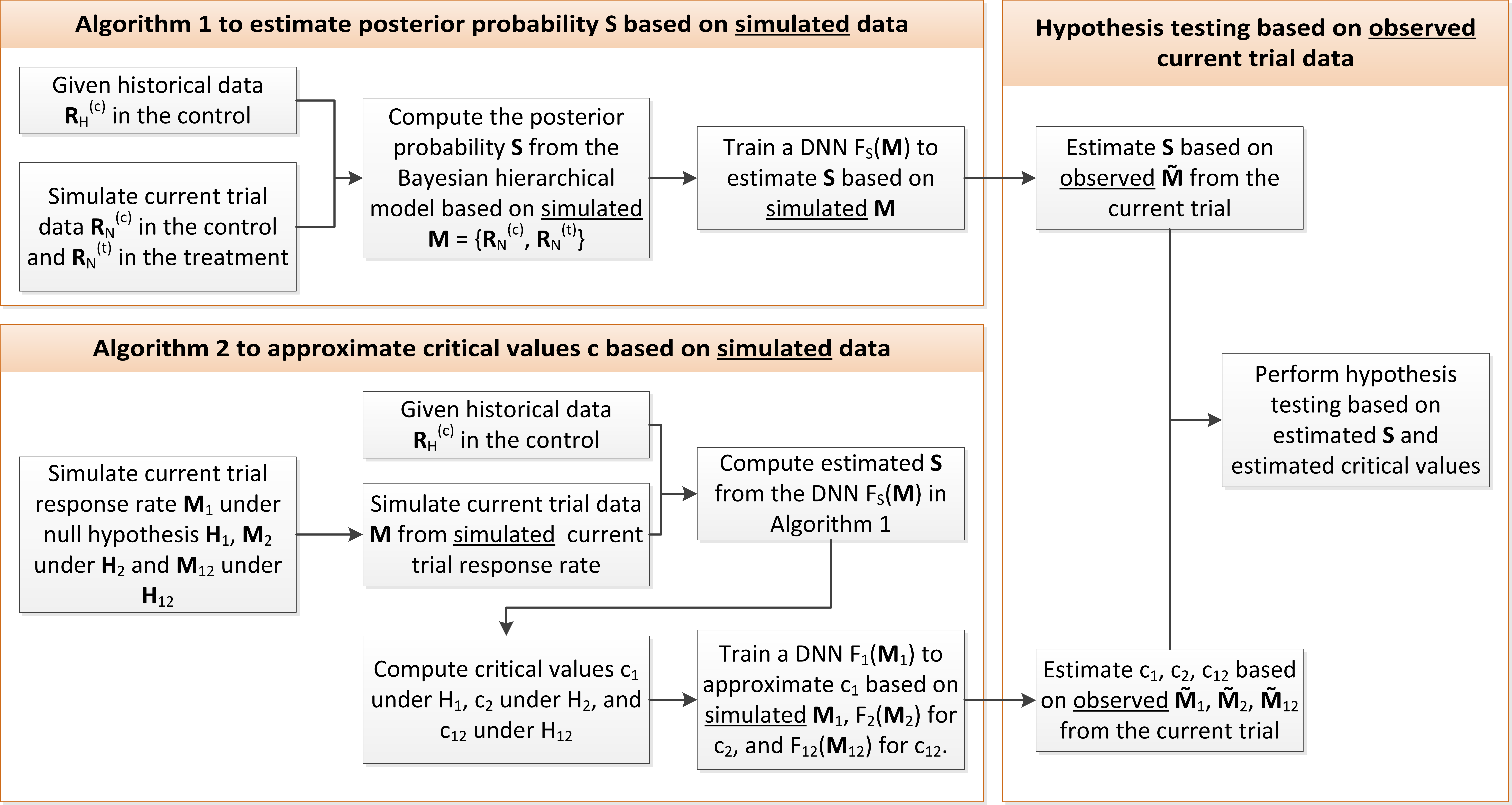}
	\label{F:chart}
\end{figure}

In Figure \ref{F:chart}, we provide a flowchart to illustrate the role of Algorithms 1 and 2 based on simulated current trial data, and hypothesis testing based on observed current trial data. 

\section{Simulation studies}
\label{sec:sim}

In this section, we conduct simulation studies to evaluate the performance of our proposed method, and compare it with the Meta-Analytic-Prior (MAP) approach
\citep{neuenschwander2010summarizing, schmidli2014robust}. Suppose that there are two endpoints ($I=2$) to be evaluated in a randomized clinical trial comparing a treatment versus placebo with equal sample size $n_0^{(c)} = n_0^{(t)} = 150$. The control information is also available in $J=6$ historical studies with sample sizes $100, 100, 200, 200, 300$, and $300$, with the response rates $0.4$ and $0.3$ for the two endpoints. We consider FWER to be controlled at $\alpha = 0.05$. 

We consider historical control data as in Table \ref{sim:table_hist_data_ind}. The number of responders $R^{(c)}_{i,j}$ for endpoint $i$, $i=1, 2$, study $j$, $j = 1, \cdots, 6$ is simulated from a Binomial distribution with rate $0.4$ for endpoint $i=1$ and $0.3$ for endpoint $i=2$. The empirical correlation of estimated response rate $R^{(c)}_{i,j}/n^{(c)}_j$ between two endpoints is $0.01$. Additional simulation studies with an empirical correlation around $0.5$ are conducted in the Supplemental Materials and demonstrate consistent findings. 

\begin{table}[ht]
\centering
\begin{tabular}{ccccccc}
\toprule
$j$ & 1 & 2 & 3 & 4 & 5 & 6 \\
\midrule
$n^{(c)}_j$ &  100 & 100 & 200 & 200 & 300 & 300 \\[0.2cm]
$R^{(c)}_{1,j}$ & 33 &  41 &  78 &  81 & 115 & 113 \\ [0.2cm]
$R^{(c)}_{2,j}$ &  31 &  28 &  69 &  68 &  94 &  97 \\ 
\bottomrule
\end{tabular}
\caption{Control data from $J=6$ historical studies.}
\label{sim:table_hist_data_ind}
\end{table}

To implement our DNN-based method, we first approximate the posterior probability $\boldsymbol{S}$ in (\ref{equ:bay_interest}) based on Algorithm \ref{alg_approx} with training data size $B = 8,000$. Its training input data $\boldsymbol{M} = \left\{ {R}^{(c)}_{1, 0}, {R}^{(c)}_{2, 0}, {R}^{(t)}_{1, 0}, {R}^{(t)}_{2, 0} \right\}$ are drawn from Binomial distributions with rates $\left\{ {\psi}^{(c)}_{1, 0}, {\psi}^{(c)}_{2, 0}, {\psi}^{(t)}_{1, 0}, {\psi}^{(t)}_{2, 0} \right\}$, which are further simulated from the following $4$ patterns with equal size $B/4 = 2,000$,
\begin{enumerate}
\item ${\psi}^{(c)}_{1, 0} \sim \mathrm{Unif}(0.2, 0.7); \:\:\: {\psi}^{(c)}_{2, 0} \sim \mathrm{Unif}(0.1, 0.6); \:\:\: {\Delta}_{1, 0}=0;  \:\:\: {\Delta}_{2, 0} = 0 $,
\item ${\psi}^{(c)}_{1, 0} \sim \mathrm{Unif}(0.2, 0.7); \:\:\: {\psi}^{(c)}_{2, 0} \sim \mathrm{Unif}(0.1, 0.6); \:\:\: {\Delta}_{1, 0}\sim \mathrm{Unif}(-0.1, 0.2);  \:\:\: {\Delta}_{2, 0} = 0 $,
\item ${\psi}^{(c)}_{1, 0} \sim \mathrm{Unif}(0.2, 0.7); \:\:\: {\psi}^{(c)}_{2, 0} \sim \mathrm{Unif}(0.1, 0.6); \:\:\: {\Delta}_{1, 0}=0;  \:\:\: {\Delta}_{2, 0} \sim \mathrm{Unif}(-0.1, 0.2) $,
\item ${\psi}^{(c)}_{1, 0} \sim \mathrm{Unif}(0.2, 0.7); \:\:\: {\psi}^{(c)}_{2, 0} \sim \mathrm{Unif}(0.1, 0.6); \:\:\: {\Delta}_{1, 0}\sim \mathrm{Unif}(-0.1, 0.2);  \:\:\: {\Delta}_{2, 0} \sim \mathrm{Unif}(-0.1, 0.2) $,
\end{enumerate}
where $\mathrm{Unif}$ denotes the Uniform distribution, and the treatment response rates are calculated at ${\psi}^{(t)}_{i, 0} = {\psi}^{(c)}_{i, 0} + {\Delta}_{i, 0}$ for $i = 1, 2$. The ranges of above Uniform distributions are determined based on previous knowledge that ${\psi}^{(c)}_{1, 0}$ is between $0.2$ and $0.7$, while ${\psi}^{(c)}_{2, 0}$ is within $0.1$ and $0.6$. The Uniform distribution can be substituted by other sampling distributions, e.g., flat normal distribution. For this simulation study, the training data size $B=8,000$ is sufficient to give DNN satisfactory performance, for example training MSE less than $10^{-3}$. It can be increased to accommodate more distinct training features, as in the next section of case study.

Next we obtain posterior samples of ${\psi}_{i, 0}^{(c)}$ from model (\ref{equ:bay_hie_model}) based on the Markov chain Monte Carlo (MCMC) method implemented by the R package {\texttt{R2jags}} \citep{su2015package}. We put a vague prior on $\boldsymbol{\theta}$ following a Normal distribution with mean zero and precision $0.01$, and an $\mathrm{InverseWishart}(\Sigma_0, k)$ prior for $\Sigma$ with $\Sigma_0$ as a unit diagonal matrix and $k = I+1$. The convergence of the MCMC algorithm is checked by the criteria of $\widehat{R}<1.01$ among $3$ chains, where $\widehat{R}$ is the ratio of between-chain versus within-chain variability \citep{gelman1992inference, berry2010bayesian}. The posterior distribution of ${\psi}_{i, 0}^{(t)}$ is a Beta distribution with a non-informative Beta prior $a_i = b_i = 1$ based on the Beta-Binomial conjugate model in (\ref{equ:trt_model}). Our posterior probability $S_i$ in (\ref{equ:bay_interest}) is evaluated by $30,000$ posterior samples with $\theta_i = 0$. In Step 2 of selecting a proper DNN structure, we consider a pool of four candidate structures:  $H_l = 2$ hidden layers with $H_n = 40$ nodes per layer, $H_l = 2$ and $H_n = 60$, $H_l = 3$ and $H_n = 40$, $H_l = 3$ and $H_n = 60$. The batch size $H_b$ is $100$, and the number of training epochs $H_e$ is $1,000$ with dropout rate $H_d = 0.1$. The above parameters are used throughout this article unless specified otherwise. In this simulation study, we choose a DNN structure with $H_l = 2$ hidden layers, $H_n = 60$ nodes per layer with the smallest validation MSE among four candidates. This DNN structure is utilized in Algorithm 2 as well. From the final fitting at Step 3, we obtain DNN $\widehat{\boldsymbol{S}} = F_S(\boldsymbol{M}; \widehat{\boldsymbol{\phi}})$ to estimate $\boldsymbol{S}$ in (\ref{equ:bay_interest}). 

In Algorithm \ref{alg_critical} of approximating the critical values, we simulate $B_1 = B_2 = B_{12} = 2,000$ datasets to reflect patterns of $H_1$ in (\ref{def:error_H1}), $H_2$ in (\ref{def:error_H2}) and $H_{12}$ in (\ref{def:error_H12}),
\begin{enumerate}
\item ${\psi}^{(c, t)}_{1, 0} \sim \mathrm{Unif}(0.2, 0.7); \:\:\: {\psi}^{(c)}_{2, 0} \sim  \mathrm{Unif}(0.1, 0.6); \:\:\: {\Delta}_{2, 0} \sim  \mathrm{Unif}(-0.1, 0.2); \:\:\: {\psi}^{(t)}_{2, 0} = {\psi}^{(c)}_{2, 0} + {\Delta}_{2, 0} $,
\item ${\psi}^{(c)}_{1, 0} \sim Unif(0.2, 0.7); \:\:\:  {\Delta}_{1, 0} \sim  \mathrm{Unif}(-0.1, 0.2); \:\:\: {\psi}^{(t)}_{1, 0} = {\psi}^{(c)}_{1, 0} + {\Delta}_{1, 0} ; \:\:\: {\psi}^{(c, t)}_{2, 0} \sim  \mathrm{Unif}(0.1, 0.6) $,
\item ${\psi}^{(c, t)}_{1, 0} \sim  \mathrm{Unif}(0.2, 0.7); \:\:\: {\psi}^{(c, t)}_{2, 0} \sim  \mathrm{Unif}(0.1, 0.6) $.
\end{enumerate}
The number of iterations of calculating critical values are $B^\prime_1 = B^\prime_2 = B^\prime_{12} = 100,000$. 

We implement the Meta-Analytic-Predictive (MAP) priors
\citep{neuenschwander2010summarizing} and two robust MAP priors with a weight of $w=50\%$ and $w=80\%$ non-informative component \citep{schmidli2014robust} by the R package {\texttt{RBesT}} \citep{rbest}. These methods handle data from each endpoint separately, instead of modeling them jointly as in our model (\ref{equ:bay_hie_model}). The setup follows their default settings with a weakly informative {Half-Normal} $(0, 1)$ prior on the standard deviation of the logit of the response rate \citep{weber2019applying}. Hypothesis testing is also based on posterior probabilities $S_i$, but the constant critical values $\widetilde{c}$ are chosen by a grid search method to control testing type I error rates not exceeding $\alpha = 0.05$ within a certain range of null response rates in the following Table \ref{sim:table_error_power}.

In Table \ref{sim:table_error_power}, we first evaluate the error rates of falsely rejecting $H_{12}$, $H_1$ or $H_2$ under the global null hypothesis where $\Delta_{1, 0} = \Delta_{2, 0} = 0$. The number of iterations in testing is $100,000$. Our proposed method has relatively accurate control on three error rates at $\alpha = 0.05$ across three scenarios with varying $\psi_{1, 0}^{(c)}$ and $\psi_{2, 0}^{(c)}$. For MAP and two robust MAPs, we choose their constant critical values $\tilde{c}$'s (introduced in Section \ref{sec:bay}) at $0.9977$, $0.9927$, and $0.9867$, respectively, by the grid search method such that the probability of rejecting $H_{12}$ reaches the nominal level of $0.05$ under the ``worst case scenario'' with $\psi_{1, 0}^{(c)} = 0.5$ and $\psi_{2, 0}^{(c)} = 0.4$. This scenario has the largest type I error rate given the same critical value under the null space evaluated in Table \ref{sim:table_error_power}. This simulation-based type I error rate control ensures that FWER does not exceed $\alpha$ under all three null cases, but leads to conservative error rates and potential power loss in some cases as evaluated later. The critical value $\tilde{c} = 0.9977$ of MAP is interpreted as a cutoff to claim a significant treatment effect based on endpoint $i$, because we reject the null hypothesis pertaining to endpoint $i$ if its posterior probability $S_i$ in (\ref{equ:bay_interest}) is larger than $\tilde{c}$ based on Section \ref{sec:bay}. 
 
Under a single null hypothesis where only a single $\Delta_i$ is equal to zero, the error rate happens when this particular true null hypothesis is erroneously rejected. All methods control this error rate well below $\alpha$. When it comes to alternative hypotheses, our method has a much higher power of rejecting each elementary null hypothesis, and a higher power of rejecting at least one of them than MAP and two RMAPs under $\psi_{1, 0}^{(c)} = 0.3, \psi_{2, 0}^{(c)} = 0.2$ and $\psi_{1, 0}^{(c)} = 0.4, \psi_{2, 0}^{(c)} = 0.3$. This is mainly due to the conservative type I error of using a constant critical value for MAP and RMAPs. When response rates are higher at $\psi_{1, 0}^{(c)} = 0.5, \psi_{2, 0}^{(c)} = 0.4$, MAP usually has the best power performance, followed by our DNN method, and then two RMAPs. Note that there are two scenarios with prior-data conflict given that historical control rates are $0.4$, $0.3$ for the first and the second endpoint, respectively. 

Table \ref{sim:table_bias} presents the bias and Table \ref{sim:table_mse} shows the root of mean squared error (RMSE) of posterior means of $\psi_{1, 0}^{(c)}$ and $\psi_{2, 0}^{(c)}$. In scenarios where the current control rates and historical rates are consistent at $0.4$ for the first endpoint and $0.3$ for the second endpoint, all methods have small biases, and MAP has the smallest RMSE. Under cases with prior-data conflict, DNN and two RMAPs have better bias than MAP, and moreover DNN has the smallest RMSE. The overall conclusion is that our DNN-based method enjoys the robustness of RMAP under prior-data conflict, and preserves power by modeling critical values as compared with MAP and RMAP by using constant critical values. 

\begin{figure}[H]
\centering
\includegraphics[scale=0.21]{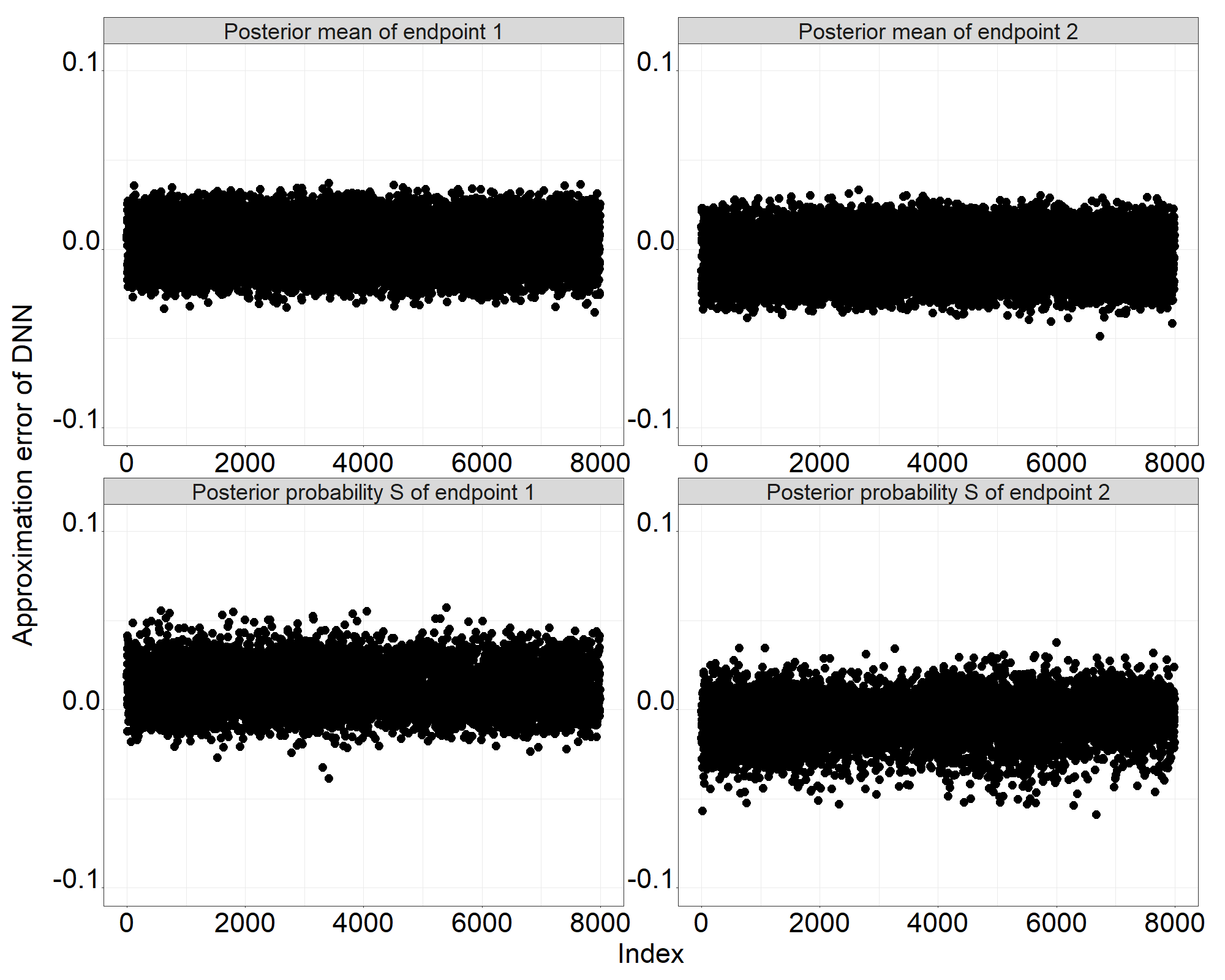}
\caption{Approximation error of DNN in estimating posterior means of $\psi_{1, 0}^{(c)}$ and $\psi_{2, 0}^{(c)}$, and posterior probabilities $S_1$ and $S_2$ in (\ref{equ:bay_interest}).}
\label{F:sim_error}
\end{figure}

Figure \ref{F:sim_error} shows the approximation errors of DNN in estimating posterior means of $\psi_{1, 0}^{(c)}$ and $\psi_{2, 0}^{(c)}$, and posterior probabilities $S_1$ and $S_2$ in (\ref{equ:bay_interest}) from $B = 8,000$ training data. The MSE from DNN training is approximately $0.001$. Errors are relatively larger in approximating $S_1$ and $S_2$ from (\ref{equ:bay_interest}), because their training labels have more randomness in the Monte Carlo estimates as compared with the empirical posterior means of $\psi^{(c)}_{1, 0}$ and $\psi^{(c)}_{2, 0}$. 

\begin{table}[ht]
	\centering
	\tiny
	\begin{tabular}{ccccccccccccccccccc}
\toprule
\multicolumn{4}{c}{} & \multicolumn{3}{c}{DNN} && \multicolumn{3}{c}{MAP} && \multicolumn{3}{c}{RMAP with $w=50\%$} && \multicolumn{3}{c}{RMAP with $w=80\%$} \\ 
 \cmidrule{5-7} \cmidrule{9-11} \cmidrule{13-15} \cmidrule{17-19}
$\psi_{1, 0}^{(c)}$ & $\psi_{2, 0}^{(c)}$ & $\Delta_{1, 0}$ & $\Delta_{2, 0}$ & $H_{12}$ & $H_1$ & $H_2$ && $H_{12}$ & $H_1$ & $H_2$ && $H_{12}$ & $H_1$ & $H_2$ && $H_{12}$ & $H_1$ & $H_2$\\ 
\midrule
\multicolumn{6}{l}{\bf Global null hypothesis} \\
\underline{0.3} & \underline{0.2} & 0 & 0 & 4.9\% & 3.0\% & 1.9\% &  & $<$0.1\% & $<$0.1\% & $<$0.1\% &  & 0.9\% & 0.3\% & 0.6\% &  & 2.1\% & 0.9\% & 1.2\% \\ 
0.4 & 0.3 &  &  & 4.9\% & 3.2\% & 1.8\% &  & 0.3\% & 0.2\% & $<$0.1\% &  & 0.9\% & 0.7\% & 0.2\% &  & 1.9\% & 1.2\% & 0.7\% \\ 
\underline{0.5} & \underline{0.4} &  &  & 4.8\% & 3.7\% & 1.2\% &  & 4.8\% & 2.6\% & 2.3\% &  & 5.1\% & 2.4\% & 2.7\% &  & 5.0\% & 2.3\% & 2.7\% \\ 
\\
\multicolumn{6}{l}{\bf Single null hypothesis} 
\\
0.4 & 0.3 & 0.1 &  & 49.3\% & 48.3\% & 1.9\% &  & 29.4\% & 29.4\% &$<$0.1\% &  & 37.7\% & 37.6\% & 0.2\% &  & 40.8\% & 40.4\% & 0.7\% \\ 
 &  & 0 & 0.1 & 42.2\% & 2.5\% & 40.9\% &  & 16.2\% & 0.2\% & 16.0\% &  & 28.1\% & 0.7\% & 27.5\% &  & 36.6\% & 1.2\% & 35.8\% \\ 
\\
\multicolumn{6}{l}{\bf Alternative hypothesis} 
\\
\underline{0.3} & \underline{0.2} & 0.1 & 0.1 & 71.8\% & 45.1\% & 47.9\% &  & 6.3\% & 3.0\% & 3.4\% &  & 30.9\% & 13.0\% & 20.6\% &  & 50.6\% & 24.8\% & 34.3\% \\ 
 &  & 0.12 & 0.12 & 85.1\% & 58.6\% & 63.0\% &  & 13.6\% & 7.1\% & 6.9\% &  & 46.1\% & 22.3\% & 30.6\% &  & 67.0\% & 37.3\% & 47.3\% \\ 
\\
0.4 & 0.3 & 0.1 & 0.1 & 67.6\% & 44.3\% & 42.8\% &  & 41.1\% & 29.6\% & 16.2\% &  & 55.0\% & 37.9\% & 27.5\% &  & 61.5\% & 40.4\% & 35.4\% \\ 
 &  & 0.12 & 0.12 & 81.3\% & 57.7\% & 57.5\% &  & 62.6\% & 46.3\% & 30.5\% &  & 74.3\% & 53.6\% & 44.5\% &  & 78.6\% & 55.1\% & 52.2\% \\ 
\\
\underline{0.5} & \underline{0.4} & 0.1 & 0.1 & 64.0\% & 45.1\% & 37.0\% &  & 68.0\% & 42.0\% & 44.8\% &  & 59.1\% & 32.6\% & 39.2\% &  & 60.7\% & 35.5\% & 39.1\% \\ 
 &  & 0.12 & 0.12 & 78.6\% & 58.1\% & 52.6\% &  & 83.3\% & 57.4\% & 60.9\% &  & 72.6\% & 44.1\% & 50.9\% &  & 75.0\% & 48.5\% & 51.5\% \\ 
\bottomrule
\end{tabular}
\begin{flushleft}
\footnotesize{Note: underlined scenarios are with prior-data conflict where historical control rates are $0.4$, $0.3$ for the first and the second endpoint, respectively. }
\end{flushleft}
\caption{Type I error rate and power of DNN-based approach, MAP and RMAP.}
\label{sim:table_error_power}
\end{table}

\begin{table}[ht]
\centering
\footnotesize
\begin{tabular}{cccccccccccccccc}
\toprule
\multicolumn{5}{c}{} & \multicolumn{2}{c}{DNN} && \multicolumn{2}{c}{MAP} && \multicolumn{2}{c}{RMAP with $w=50\%$} && \multicolumn{2}{c}{RMAP with $w=80\%$}\\ 
\cmidrule{6-7} \cmidrule{9-10} \cmidrule{12-13} \cmidrule{15-16} 
$\psi_{1, 0}^{(c)}$ & $\psi_{2, 0}^{(c)}$ & $\Delta_{1, 0}$ & $\Delta_{2, 0}$ && $\psi_{1, 0}^{(c)}$ & $\psi_{2, 0}^{(c)}$ && $\psi_{1, 0}^{(c)}$ & $\psi_{2, 0}^{(c)}$ && $\psi_{1, 0}^{(c)}$ & $\psi_{2, 0}^{(c)}$ && $\psi_{1, 0}^{(c)}$ & $\psi_{2, 0}^{(c)}$  \\ 
\midrule
\multicolumn{6}{l}{\bf Global null hypothesis} 
\\
\underline{0.3} & \underline{0.2} & 0 & 0 &  & 0.025 & 0.029 &  & 0.048 & 0.050 &  & 0.027 & 0.018 &  & 0.015 & 0.010 \\ 
0.4 & 0.3 &  &  &  & 0.004 & 0.006 &  & -0.010 & 0.015 &  & -0.007 & 0.012 &  & -0.004 & 0.008 \\ 
\underline{0.5} & \underline{0.4} &  &  &  & -0.015 & -0.016 &  & -0.050 & -0.046 &  & -0.020 & -0.024 &  & -0.009 & -0.012 \\ 
\\
\multicolumn{6}{l}{\bf Single null hypothesis} 
\\
0.4 & 0.3 & 0.1 & 0 &  & 0.004 & 0.006 &  & -0.011 & 0.015 &  & -0.008 & 0.012 &  & -0.004 & 0.008 \\ 
 &  & 0 & 0.1 &  & 0.004 & 0.006 &  & -0.011 & 0.014 &  & -0.008 & 0.011 &  & -0.004 & 0.008 \\ 
\\
\multicolumn{6}{l}{\bf Alternative hypothesis} 
\\
\underline{0.3} & \underline{0.2} & 0.1 & 0.1 &  & 0.025 & 0.029 &  & 0.046 & 0.049 &  & 0.026 & 0.018 &  & 0.015 & 0.010 \\ 
 &  & 0.12 & 0.12 &  & 0.025 & 0.029 &  & 0.048 & 0.048 &  & 0.026 & 0.018 &  & 0.014 & 0.010 \\ 
\\
0.4 & 0.3 & 0.1 & 0.1 &  & 0.004 & 0.006 &  & -0.011 & 0.015 &  & -0.008 & 0.012 &  & -0.004 & 0.008 \\ 
 &  & 0.12 & 0.12 &  & 0.004 & 0.006 &  & -0.010 & 0.015 &  & -0.007 & 0.012 &  & -0.004 & 0.008 \\ 
\\
\underline{0.5} & \underline{0.4} & 0.1 & 0.1 &  & -0.015 & -0.016 &  & -0.052 & -0.043 &  & -0.020 & -0.024 &  & -0.009 & -0.012 \\ 
 &  & 0.12 & 0.12 &  & -0.015 & -0.016 &  & -0.054 & -0.044 &  & -0.021 & -0.024 &  & -0.009 & -0.012 \\ 
\bottomrule
\end{tabular}
\begin{flushleft}
	\footnotesize{Note: underlined scenarios are with prior-data conflict where historical control rates are $0.4$, $0.3$ for the first and the second endpoint, respectively. }
\end{flushleft}
\caption{Bias of posterior means $\psi_{1, 0}^{(c)}$ and $\psi_{2, 0}^{(c)}$ in DNN, MAP and RMAP.}
\label{sim:table_bias}
\end{table}

\begin{table}[ht]
\centering
\footnotesize
\begin{tabular}{cccccccccccccccc}
\toprule
\multicolumn{5}{c}{} & \multicolumn{2}{c}{DNN} && \multicolumn{2}{c}{MAP} && \multicolumn{2}{c}{RMAP with $w=50\%$} && \multicolumn{2}{c}{RMAP with $w=80\%$}\\ 
\cmidrule{6-7} \cmidrule{9-10} \cmidrule{12-13} \cmidrule{15-16} 
$\psi_{1, 0}^{(c)}$ & $\psi_{2, 0}^{(c)}$ & $\Delta_{1, 0}$ & $\Delta_{2, 0}$ && $\psi_{1, 0}^{(c)}$ & $\psi_{2, 0}^{(c)}$ && $\psi_{1, 0}^{(c)}$ & $\psi_{2, 0}^{(c)}$ && $\psi_{1, 0}^{(c)}$ & $\psi_{2, 0}^{(c)}$ && $\psi_{1, 0}^{(c)}$ & $\psi_{2, 0}^{(c)}$  \\ 
\midrule
\multicolumn{6}{l}{\bf Global null hypothesis} 
\\
\underline{0.3} & \underline{0.2} & 0 & 0 &  & 0.039 & 0.039 &  & 0.055 & 0.061 &  & 0.048 & 0.046 &  & 0.043 & 0.038 \\ 
0.4 & 0.3 &  &  &  & 0.031 & 0.029 &  & 0.018 & 0.020 &  & 0.023 & 0.024 &  & 0.029 & 0.029 \\ 
\underline{0.5} & \underline{0.4} &  &  &  & 0.037 & 0.036 &  & 0.064 & 0.052 &  & 0.054 & 0.048 &  & 0.047 & 0.044 \\ 
\\
\multicolumn{6}{l}{\bf Single null hypothesis} 
\\
0.4 & 0.3 & 0.1 & 0 &  & 0.031 & 0.029 &  & 0.018 & 0.020 &  & 0.024 & 0.024 &  & 0.030 & 0.029 \\ 
 &  & 0 & 0.1 &  & 0.031 & 0.029 &  & 0.018 & 0.020 &  & 0.023 & 0.024 &  & 0.030 & 0.029 \\ 
\\
\multicolumn{6}{l}{\bf Alternative hypothesis} 
\\
\underline{0.3} & \underline{0.2} & 0.1 & 0.1 &  & 0.039 & 0.039 &  & 0.053 & 0.062 &  & 0.048 & 0.046 &  & 0.043 & 0.038 \\ 
 &  & 0.12 & 0.12 &  & 0.039 & 0.039 &  & 0.054 & 0.060 &  & 0.048 & 0.045 &  & 0.043 & 0.038 \\ 
\\
0.4 & 0.3 & 0.1 & 0.1 &  & 0.031 & 0.029 &  & 0.018 & 0.020 &  & 0.024 & 0.024 &  & 0.030 & 0.029 \\ 
 &  & 0.12 & 0.12 &  & 0.031 & 0.029 &  & 0.018 & 0.020 &  & 0.023 & 0.024 &  & 0.029 & 0.028 \\ 
\\
\underline{0.5} & \underline{0.4} & 0.1 & 0.1 &  & 0.037 & 0.036 &  & 0.066 & 0.052 &  & 0.054 & 0.048 &  & 0.047 & 0.044 \\ 
 &  & 0.12 & 0.12 &  & 0.037 & 0.036 &  & 0.065 & 0.051 &  & 0.054 & 0.047 &  & 0.047 & 0.044 \\  
\bottomrule
\end{tabular}
\begin{flushleft}
	\footnotesize{Note: underlined scenarios are with prior-data conflict where historical control rates are $0.4$, $0.3$ for the first and the second endpoint, respectively. }
\end{flushleft}
\caption{RMSE of posterior means $\psi_{1, 0}^{(c)}$ and $\psi_{2, 0}^{(c)}$ in DNN, MAP and RMAP.}
\label{sim:table_mse}
\end{table}

\section{A case study}
\label{sec:case}

We design a generic randomized clinical trial evaluating the efficacy of a study drug versus an active comparator secukinumab $300$ mg \citep{langley2014secukinumab} in patients with moderate-to-severe plaque psoriasis with equal sample size per group $n_0^{(c)} = n_0^{(t)} = 200$. We consider the co-primary endpoints in \citet{langley2014secukinumab}: the proportion of patients achieving a reduction of $75\%$ or more from baseline in the Psoriasis Area-and-Severity Index score (PASI 75) and the proportion of patients achieving a score of 0 (clear) or 1 (almost clear) on a 5-point modified investigator’s global assessment (MIGA 0/1) at week 12. The control information is available in $J=3$ historical studies: ERASURE, FIXTURE \citep{langley2014secukinumab} and JUNCTURE \citep{paul2015efficacy} with data summarized in Table \ref{case:table_hist_data_ind}. The weighted observed response rates are approximately $0.80$ and $0.65$ for PASI 75 and MIGA 0/1, respectively. We evaluate the performance of different methods on the following three scenarios on response rates from the current trial:
\begin{enumerate}
\item Prior-data conflict scenario 1 (S1): ${\psi}^{(c)}_{1, 0} = 0.7$ and ${\psi}^{(c)}_{2, 0} = 0.55$,
\item Prior-data conflict scenario 2 (S2): ${\psi}^{(c)}_{1, 0} = 0.9$ and ${\psi}^{(c)}_{2, 0} = 0.75$,
\item Prior-data consistent scenario (S3): ${\psi}^{(c)}_{1, 0} = 0.8$ and ${\psi}^{(c)}_{2, 0} = 0.65$.
\end{enumerate}

\begin{table}[ht]
\centering
\begin{tabular}{ccccc}
	\toprule
	Historical study & ERASURE & FIXTURE & JUNCTURE & Total  \\
	\midrule
	$n^{(c)}_j$ &  245 & 323 & 60 & 628  \\[0.2cm]
	$R^{(c)}_{1,j}$ (rate) of PASI 75 & 200 (0.82) & 249 (0.77) & 52 (0.87) & 501 (0.80) \\ [0.2cm]
	$R^{(c)}_{2,j}$ (rate) of MIGA 0/1 &  160 (0.65) & 202 (0.63) & 44 (0.73) & 406 (0.65) \\ 
	\bottomrule
\end{tabular}
\caption{Data of the active comparator secukinumab $300$ mg in $J=3$ historical studies.}
\label{case:table_hist_data_ind}
\end{table}

When generating training data in our method, we consider the range of ${\psi}^{(c)}_{1, 0}$ as $0.65$ to $0.95$, the range of ${\psi}^{(c)}_{2, 0}$ as $0.5$ to $0.8$, and the ranges of ${\Delta}_{1, 0}$ and ${\Delta}_{2, 0}$ as $-0.1$ to $0.1$. The choices of above ranges are based on the team's knowledge, such as the historical data in Table \ref{case:table_hist_data_ind}, target product profile (TPP) of the new drug, et cetera. Theses ranges can be set wider as needed. As compared with simulation studies in Section \ref{sec:sim}, we increase the training data size $B$ in Algorithm 1 from $8,000$ to $16,000$ to get DNN training MSE less than $10^{-3}$. This is to accommodate a larger number of distinct features from training input data $\boldsymbol{M}$ due to larger current sample sizes $n_0^{(c)}$ and $n_0^{(t)}$. The constant critical values in MAP, RMAP with $w=50\%$, and RMAP with $w=80\%$ are $0.987$, $0.984$, and $0.980$, respectively, to protect maximum testing type I error rates not exceeding $\alpha = 0.05$ under three scenarios S1, S2, and S3. The computation of constant critical values of MAP and RMAP is discussed in Section \ref{sec:sim}. The selected DNN has $H_l = 2$ hidden layers and $H_n = 60$ nodes per layer. Other parameter setups are the same as Section \ref{sec:sim}. 

\begin{figure}[p]
\centering
\begin{subfigure}{1\linewidth}
\centering
\caption{Power of rejecting at least one null hypothesis.}
\includegraphics[scale=0.17]{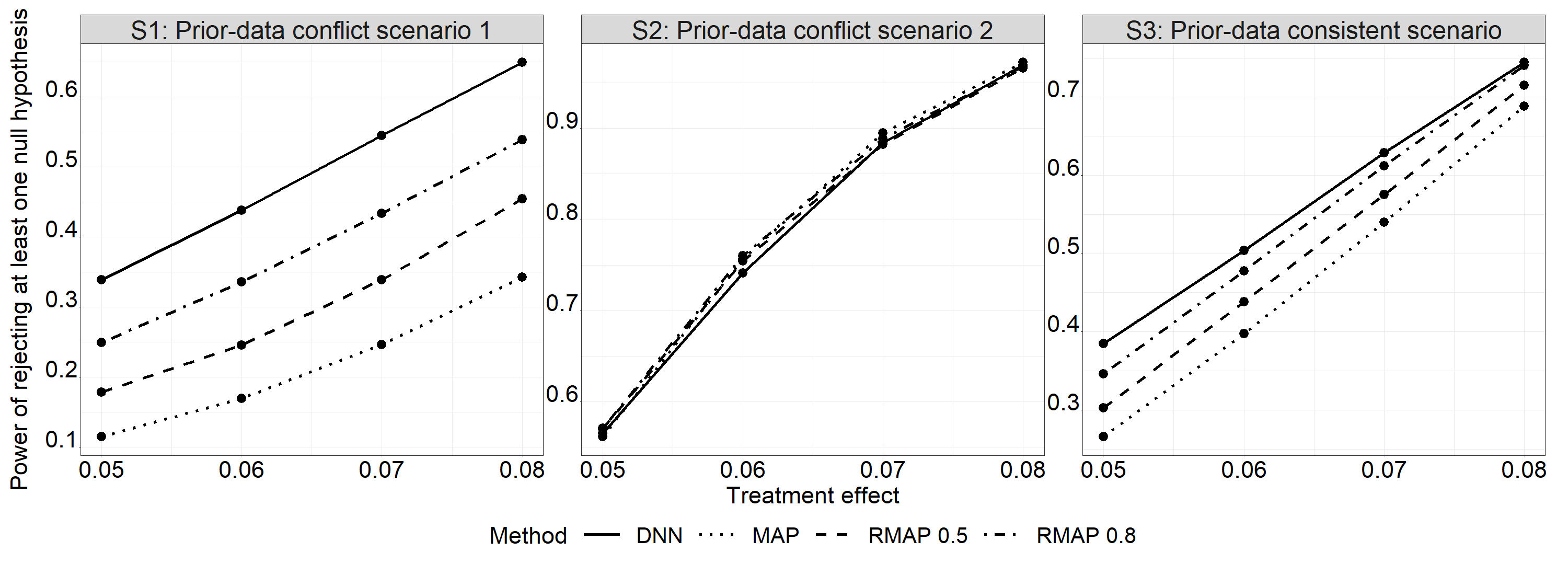}
\label{F:case_power_1}
\end{subfigure}
\hfill
\begin{subfigure}{1\linewidth}
\centering
\caption{Power of rejecting the first null hypothesis.}
\includegraphics[scale=0.17]{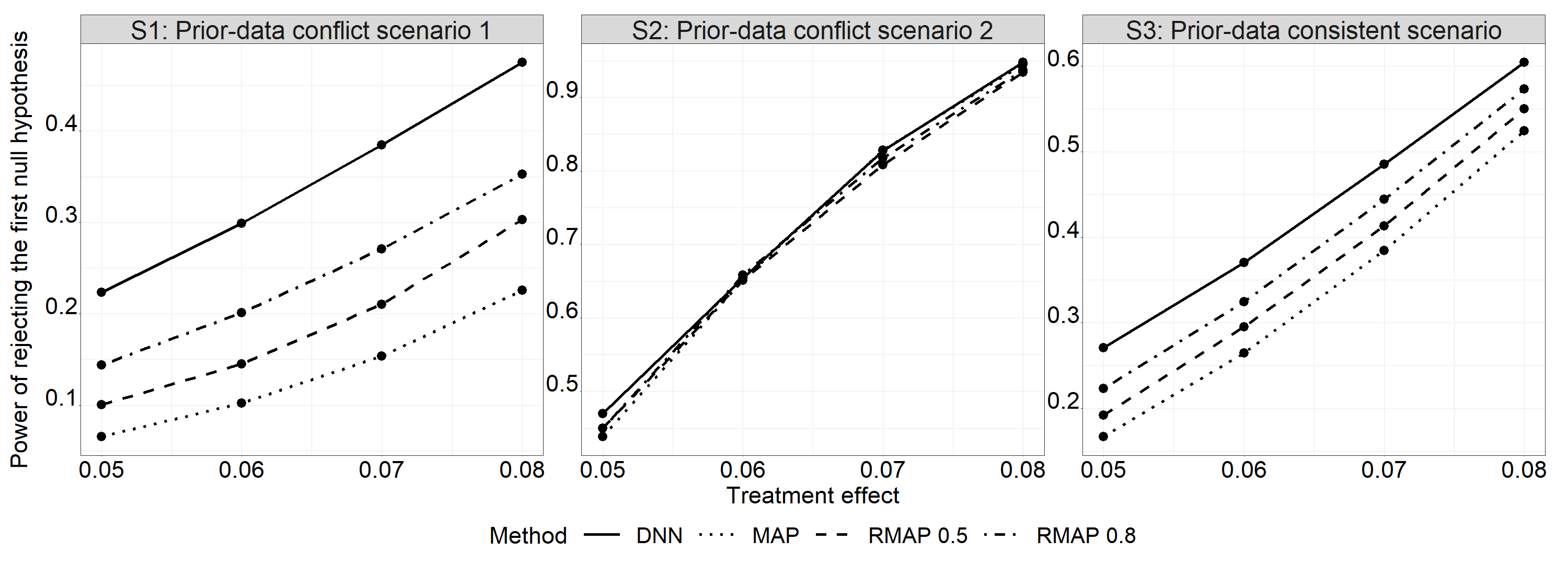}
\label{F:case_power_2}
\end{subfigure} 
\hfill
\begin{subfigure}{1\linewidth}
	\centering
	\caption{Power of rejecting the second null hypothesis.}
	\includegraphics[scale=0.17]{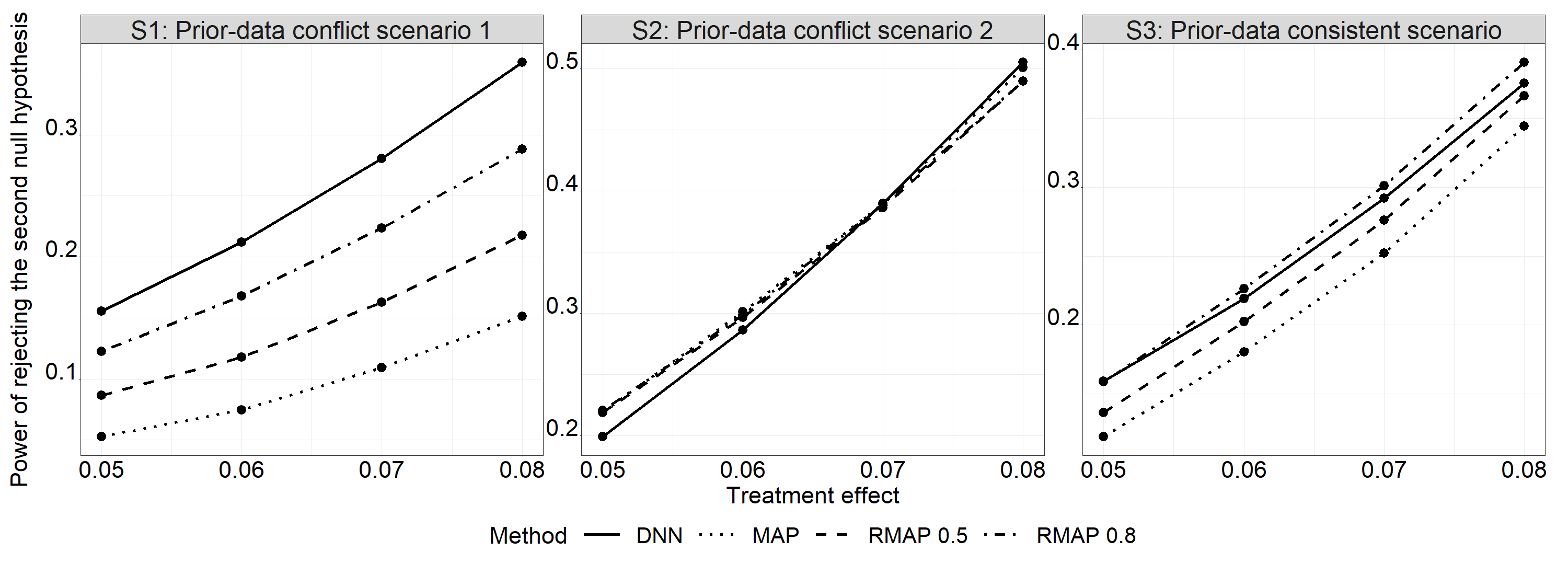}
	\label{F:case_power_3}
\end{subfigure} 
	\caption{Power performance of DNN, MAP and two RMAPs.}
	\label{F:case_power}
	
\end{figure}

\begin{figure}[p]
\centering
\begin{subfigure}{1\linewidth}
	\centering
	\caption{Absolute bias of the posterior mean $\psi_{1, 0}^{(c)}$.}
	\includegraphics[scale=0.17]{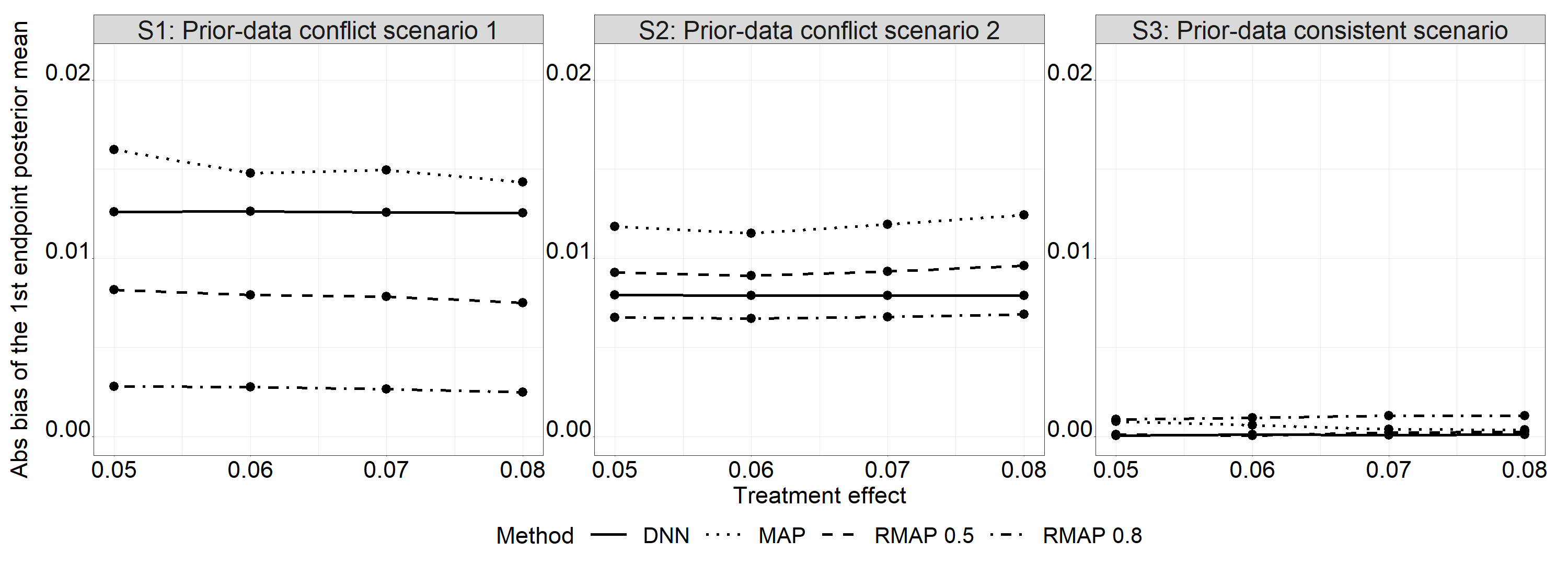}
	\label{F:case_bias_1}
\end{subfigure}
\hfill
\begin{subfigure}{1\linewidth}
	\centering
	\caption{Absolute bias of the posterior mean $\psi_{2, 0}^{(c)}$.}
	\includegraphics[scale=0.17]{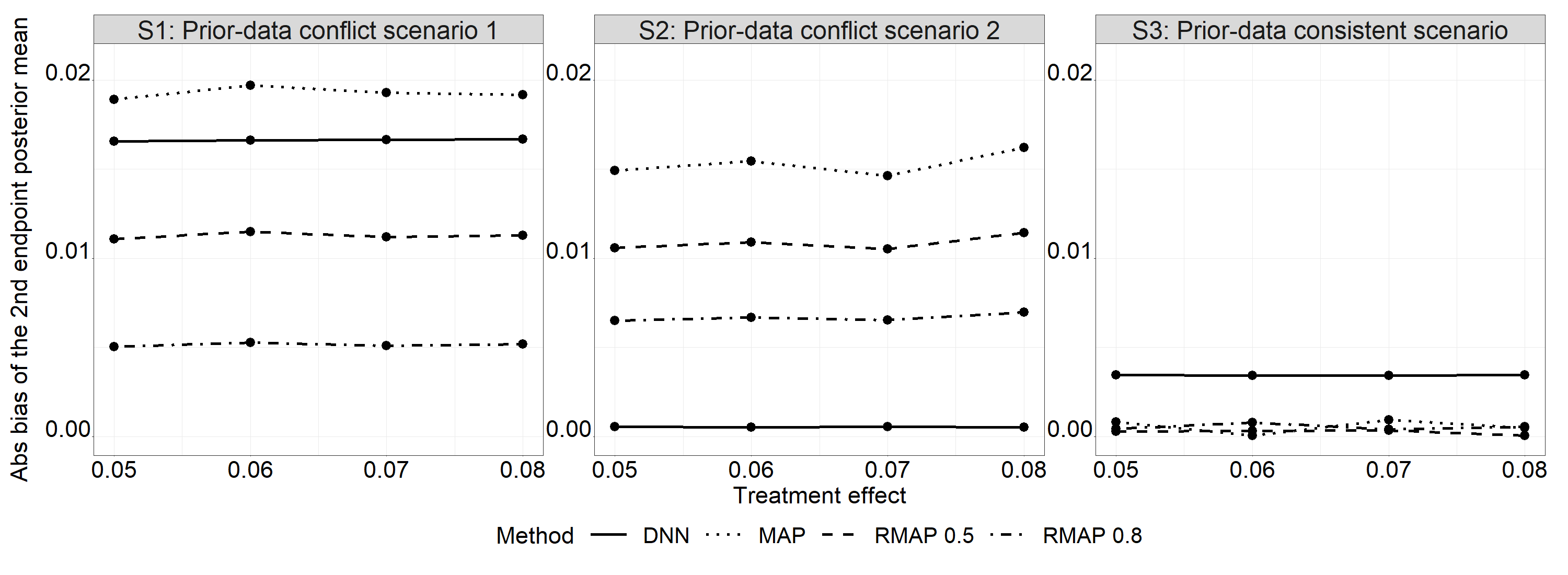}
	\label{F:case_bias_2}
\end{subfigure} 
\caption{Absolute bias of posterior means $\psi_{1, 0}^{(c)}$ and $\psi_{2, 0}^{(c)}$ in DNN, MAP and two RMAPs.}
\label{F:case_bias}
\end{figure}

\begin{figure}[p]
\centering
\begin{subfigure}{1\linewidth}
	\centering
	\caption{RMSE of the posterior mean $\psi_{1, 0}^{(c)}$.}
	\includegraphics[scale=0.17]{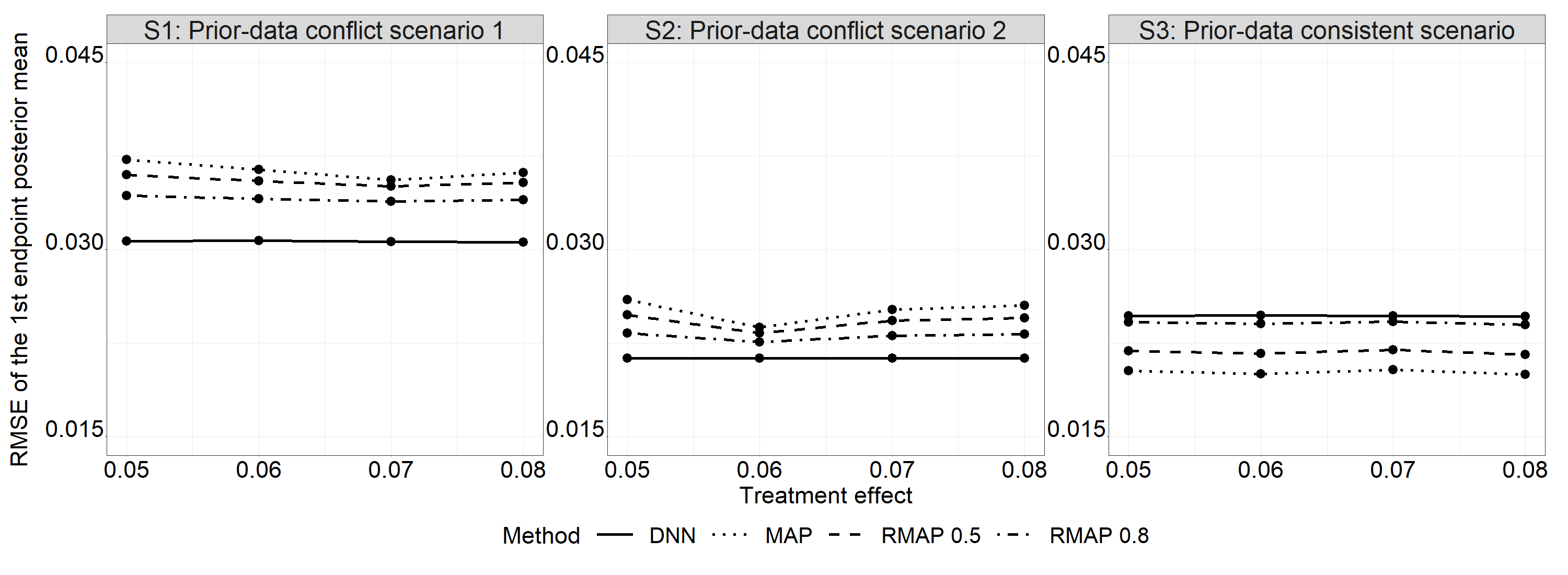}
	\label{F:case_RMSE_1}
\end{subfigure}
\hfill
\begin{subfigure}{1\linewidth}
	\centering
	\caption{RMSE of the posterior mean $\psi_{2, 0}^{(c)}$.}
	\includegraphics[scale=0.17]{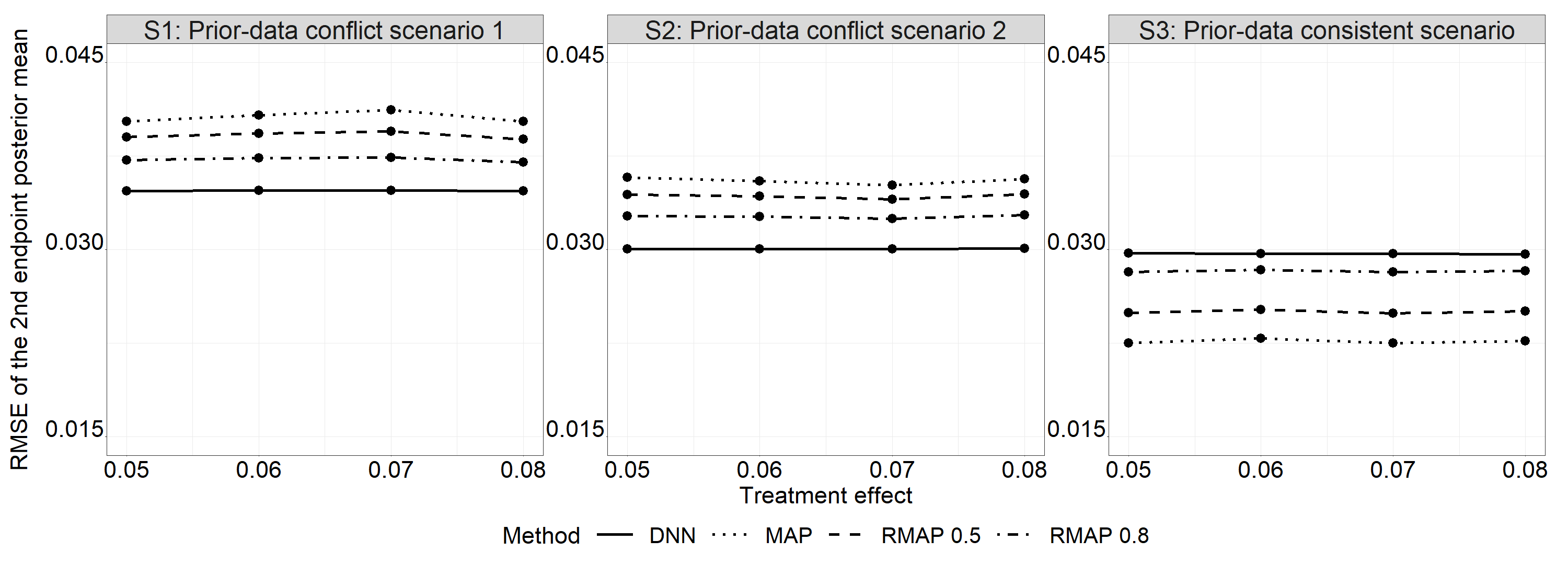}
	\label{F:case_RMSE_2}
\end{subfigure} 
\caption{RMSE of posterior means $\psi_{1, 0}^{(c)}$ and $\psi_{2, 0}^{(c)}$ in DNN, MAP and two RMAPs.}
\label{F:case_RMSE}
\end{figure}

Similar to what we observe on type I error rates in Table \ref{sim:table_error_power}, MAP and two RMAPs have conservative type I error rates under scenarios with lower response rates (i.e., S1 and S3).
In terms of power, our DNN-based method generally has a higher probability of rejecting at least one null hypothesis (Figure \ref{F:case_power_1}), rejecting the first null hypothesis (Figure \ref{F:case_power_2}), and rejecting the second null hypothesis (Figure \ref{F:case_power_3}) than MAP and two RMAPs under S1 and S3. Our method preserves power in these cases by modeling critical values in Algorithm 2. Under S2 where all methods have type I error rates of approximately $5\%$, all methods have similar power performance. The RMAP methods demonstrate the smallest absolute bias for the posterior means $\psi_{1, 0}^{(c)}$ (Figure \ref{F:case_bias_1}) and $\psi_{2, 0}^{(c)}$ (Figure \ref{F:case_bias_2}) in general. Our DNN approach has smaller bias than MAP under scenarios S1 and S2 with prior-data conflict. The RMSE of our method is the smallest under two prior-data conflict scenarios, but is slightly larger than comparators under the prior-data consistent scenario (Figure \ref{F:case_RMSE_1}, \ref{F:case_RMSE_2}). Therefore, our proposed method has satisfactory RMSE under prior-data conflicts, and preserves power by estimating critical values with DNN.

\section{Concluding remarks}
\label{sec:discussion}

In this article, we construct a prospective DNN-based framework from the Bayesian hierarchical model to synthesize control information from multiple endpoints. Our two-stage method first approximates posterior probabilities and then estimates critical values. The decision functions for hypothesis testing based on DNN training can be locked in files before initiation of the current trial to ensure study integrity, which is appealing to regulatory agencies.  

Our DNN-based prospective algorithm can also save computational time as compared with the traditional simulation method. Taking the case study in Section \ref{sec:case} for illustration, there are $12$ setups ($4$ magnitudes of treatment effect $\times$ $3$ scenarios) in total and $100,000$ testing iterations per setup. As shown in Table \ref{computation_time}, it takes approximately $34$ hours for DNN to conduct the computation, while over $700$ hours for the traditional method. The main saving is due to the fact that DNN builds an approximating function in Algorithm 1 with only $B = 16, 000$ rounds of MCMC computation required. The final testing is fast based on well-trained DNNs. On the contrary, the traditional approach needs to perform MCMC for every iteration ($1,200,000$ in total) in testing. 

\begin{table}[ht]
	\centering
	\begin{tabular}{ccccc}
		\toprule
		Method & Algorithm 1 & Algorithm 2 & Testing & Total time\\
		\midrule
		DNN & 11.89 & 22.08 & 0.04 & 34.01  \\
		Traditional Simulation & - & - & $ >700$ & $>700$   \\
		\bottomrule
	\end{tabular}
	\caption{Computational time (in hours) of DNN and MCMC in the case study.}
	\label{computation_time}
\end{table}

Another important contribution of our work is to model the critical values by DNN to control FWER. A common practice is to choose the cutoff value by a grid-search method to control type I errors in testing within a certain range of the null space. Simulations show a moderate power gain of our proposed method, especially when the constant critical value has a conservative error rate. To accommodate approximation errors, a smaller working significance level can be utilized to control validated type I error rates strictly smaller than the nominal level, if necessary. Our framework can be broadly generalized to other types of Bayesian designs when the critical value is not available analytically in finite samples. 

We discuss some potential limitations of the proposed method. Firstly, there is a lack of theoretical understanding of the upper bound of the approximation error when estimating posterior probabilities $\boldsymbol{S}$ by DNN in Algorithm 1, and estimating critical values by DNN in Algorithm 2. In this article, we provide empirical evidence to address this by checking model fitting MSE and the error plot in Figure \ref{F:sim_error}. Moreover, our method requires a few more hours in simulating training data for DNNs before the current trial conduct, as shown in Table \ref{computation_time}. However, after observing current trial data, we can instantly compute the posterior probabilities and critical values to conduct hypothesis testing. Table \ref{computation_time} shows that our method saves considerable computational time compared to the traditional simulation method in the testing stage.

\section*{Supplemental Materials}

Supplemental Materials include additional simulation results of Section \ref{sec:sim} with an empirical correlation at $0.5$. The R code is available at \url{{https://github.com/tian-yu-zhan/Deep_Historical_Borrowing}}.

\section{Acknowledgements}
This manuscript was sponsored by AbbVie, Inc. AbbVie contributed to the design, research, and interpretation of data, writing, reviewing, and approving the content. Tianyu Zhan, Ziqian Geng, Yihua Gu, Li Wang and Xiaohong Huang are employees of AbbVie Inc. Yiwang Zhou was a 2020 summer intern at AbbVie Inc. Jian Kang is Professor at Department of Biostatistics, University of Michigan, Ann Arbor. Professor Kang’s research was partially supported by NIH R01 GM124061 and R01 MH105561. Elizabeth H. Slate is Distinguished Research Professor, Duncan McLean and Pearl Levine Fairweather Professor at Department of Statistics, Florida State University. This work was initiated when Professor Slate was the AbbVie Visiting Scholar in Honor of David C. Jordan from 2018 to 2019. All authors may own AbbVie stock.

Authors would also like to thank the editor Margaret Gamalo, an anonymous associate editor and three anonymous reviewers for their constructive comments.

\bibliographystyle{apalike}
\bibliography{./HBM_ref}

\end{document}